# Voltage Profile Improvement of Distribution Grid by Using a New Control Approach on Injected Reactive Power of Plug-in Electric Vehicle Parking Lots to Grid

Mohammad Mehdi Rezvani Louisiana state university
Baton Rouge, USA
mrezva2@lsu.edu

Reza Khoud, Niroo Pars Company
Yazd, Iran
rkhoud@gmail.com

*Abstract*—**Establishing the electrical parking lots will become more important by increasing use of electric vehicles (EVs). These parking lots not only can be seen as high electrical power consumption loads which can cause the voltage drop at feeders but also they can be used as electrical power plants that can help the main power grid during the load peak hour or any congestion. Therefore, finding an optimum place for the building of these lots, in which the deviation of the voltage at the feeder, power loss in the grid, and cost be minimized, is essential.**

**In this paper, a new method of vehicle-to-grid reactive power support (V2GQ) has been used to add the model of plug-in electric vehicle (PEV) as one part of the objective function to find the optimum place of parking lots. The non-sorting genetic algorithm II (NSGA2) is used here as an optimization algorithm to find the optimum voltage profile based on the location of parking lots. For validation of the purposed method, a 33-bus standard distribution network has been studied**.

*Keywords: Smart grid, distributed generation, plug-in electric vehicles, optimal placement, parking lots of PEVs.*

## I. INTRODUCTION

Increasing in penetration of electric vehicles have some effect on the distribution network such as enhancement or decrease of the nodes' voltage. The content of V2GQ will be obtained, by using PEV for charging these kinds of vehicles. Since the voltage of the distribution system is base on radial power flow, changes in PEV's consumption and distributed generations' (DG) output power have an impact on different voltage nodes of the distribution system. The increase or decrease in nodes' voltage is inevitable. One of the most important issues in the power distribution network is to find an optimal way for voltage control [1-9]. In [4], not only the author uses the V2G system for representing voltage control methods based on exchanging active and reactive, but also expresses the management of charge and discharge for electric vehicles to control voltage. The author in [3] proposes a new way of reducing production cost by managing the charge of several electric vehicles when they are connecting to SG. This method is base on an adaptation of the power market price that varies with time and the preference of cars owners for charging their vehicles according to their priority. In [3], the author does not consider the potential of reactive power generation of PEV to reduce power losses. Due to the increasing penetration of DG in power networks, the usual methods for nodes' voltage control is inefficient. By the same token, there is a significant need for optimal control voltage in a distribution network with assuming distribution system's equipment such as on-load tap changer (OLTC), step voltage regulation, parallel capacitor, and static reactive power [6]. In [10], the purpose of the proposed scheme focusing on control voltage is to minimize the power losses and voltage deviation. Furthermore, this plan includes the optimal responses for OLTC, the output of DG and instruments using for controlling of reactive power.

In this paper, a new scheme for optimal placement of PEV by using the maximum capacity of inverters existing in this parking and the inverters that are used in DG, to generate reactive power and improve voltage profile, is proposed. Furthermore, the proposed method can encourage investors to invest in this area because of the increased profits of selling their excessive reactive power in the electric market. The characteristics of V2GQ technology that cause the V2G pales in comparison with it describe as follow. First, the battery of an electric vehicle (EV) doesn't discharge; as a corollary, this technology helps to increase battery life which is desirable for the owner of EV. Second, with generating reactive power and selling it in the electric market, the proprietors of these EVs can increase their profit.

## II. BASICS OF BI-DIRECTIONAL POWER TRANSFER

The old distribution systems are passive in which power can only flow in one direction, from the substation to load. While, modern distribution networks are an active system that can have bi-directional power transfer, due to the presence of DGs and PEVs. Fig. 1 shows two buses with two ideal voltage sources connected to each other by a specific impedance. With respect to the direction of $I_{12}$ each of these two bused can either generate active and reactive power or consume active and reactive power.

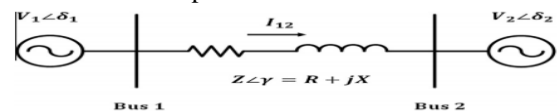

Fig. 1. A generic diagram of two interconnected buses

Active and reactive power in Fig.1 can be calculated by equations (2) and (3):

$$Z \angle \gamma = R + jX \quad V_1 \angle \delta_1 \quad V_2 \angle \delta_2 \tag{1}$$

$$P_{12} = \frac{V_1^2}{Z}\cos(\gamma) - \frac{V_1 V_2}{Z}\cos(\gamma + \delta_1 - \delta_2) \quad (2)$$

$$Q_{12} = \frac{V_1^2}{Z}\sin(\gamma) - \frac{V_1 V_2}{Z}\sin(\gamma + \delta_1 - \delta_2) \quad (3)$$

Equations (2) and (3) will be simplified to equations (4) and (5), which have only voltage magnitude and angle, if $X \gg R$, as follow:

$$P_{12} = \frac{V_1 V_2}{X}\cos(\delta_1 - \delta_2) \quad (4)$$

$$Q_{12} = \frac{V_1}{X}[V_1 - V_2 \cos(\delta_1 - \delta_2)] \quad (5)$$

Equation (4) reveals that the flow of active power is more related to the angle of voltage, while, equation (5) shows that the flow of reactive power is more affected by voltage magnitudes. The direction of flow of active and reactive power in Fig.1 is based on Table 1 [12].

Table 1. *Direction of power flow*

| Conditions | Power transfer |
|---|---|
| $\delta_1 \rangle \delta_2$ | Active power flows from bus 1 to bus 2 |
| $\delta_1 \langle \delta_2$ | Active power flows from bus 2 to bus 1 |
| $V_1 \rangle V_2$ | Reactive power flows from bus 1 to bus 2 |
| $V_1 \langle V_2$ | Reactive power flows from bus 2 to bus 1 |

III. PROBLEM DEFINITION

It is probable that the voltage profile of buses or feeders got out of range, or the in power grid experiences the voltage drop or overvoltage. For instance, overvoltage might be happened when PEVs are discharging to the power grid at the light load, and voltage drop might be occurred many PEVs in parking lots connect to the grid at peak load time to be charged. In these cases, ordinary methods such as OLTC is used for voltage regulation. The issue is becoming more significant when both overvoltage and voltage drop happen at the same time. Which means there is excessive power generation in some feeders due to the high penetration of DGs or discharging PEVs while the others are highly loaded because of charging the PEVs. In such scenarios, the OLTC confront with two opposing solutions. Decreasing the voltage ratio ($V_1/V_2$) of substations transformer will lead to improving the overvoltage problem, but it makes the voltage drop issue more severe, and vice versa. Thus, making changes in the tap of OLTC does not help to solve this issue. Toronto parking authority (TPA) [1] provides two different power profiles for conventional DG and PEV uncontrolled charging load which are depicted in Fig.2. The Fig.2 reveals the fact that the probability of having both under-voltage and overvoltage simultaneously is considerable. Therefore, the regulation of voltage must be done by considering the DGs and PEVs. Putting limitation on the generated active power of DG and supporting the reactive power of DG are two probable solutions that can be expressed by using DGs. However, restriction on the generated active power of DGs means waste of money; therefore, the first solution is not rational.

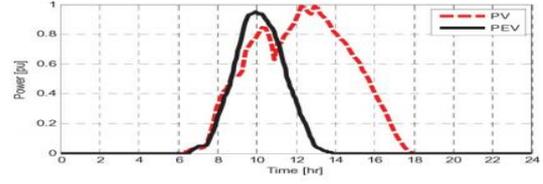

Fig. 2. DG and PEV power profiles [1]

Furthermore, the nominal power of DGs puts the inherent restraint on reactive power provision, and hence, it may not have the capability for unraveling the issue. While the extra reactive power of a plug-in electric vehicle can be exploited as a valid method to regulate voltage, one individual PEV or some scattered PEVs in the different location cannot improve the regulation of voltage well. Parking lots are a good solution to this problem because many PEVs can be connected to the same feeder at the same time. In addition, it is not economical to build parking lots at any feeders or some random feeders. Therefore, the optimization problem should be solved for this purpose. Here, the optimum placement for parking lots of PEVs is done based on the following aims. First, reducing power losses. Second, minimizing the voltage deviation from optimum voltage. Third, cut costs.

*Constraints*

It is indispensable to consider the network constraints for optimal placement of PEVs parking lots. The power flow equations, which is express in equations (6) and (7), have to be considered in all levels of optimization to satisfy constraints.

$$p_{G(i,t)} - p_{L(i,t)} = \sum_{j \in I} V_{(i,t)} V_{(i,t)} Y_{(i,j)} \cdot \cos(\theta_{(i,j)} + \delta_{(j,t)} - \delta_{(i,t)}) \quad (6)$$

$$\forall i,t$$

$$Q_{G(i,t)} - Q_{L(i,t)} = -\sum_{j \in I} V_{(i,t)} V_{(i,t)} Y_{(i,j)} \cdot \sin(\theta_{(i,j)} + \delta_{(j,t)} - \delta_{(i,t)}) \quad (7)$$

$$\forall i,t$$

where $P_{G(i,t)}$ and $Q_{G(i,t)}$ define the generated active and reactive powers at bus $i$ and time $t$, respectively; $P_{L(i,t)}$ and $Q_{L(i,t)}$ express the load active and reactive power at bus $i$ and time $t$, respectively; $V_{(i,t)}$ and $\delta_{(i,t)}$ present the magnitude and angle of the voltage at bus $i$ and time $t$, respectively; $Y_{(i,t)}$ and $\theta_{(i,t)}$ are the magnitude and angle of the Y-bus admittance matrix.

The voltage and feeder thermal restrictions ought to be held, and hence

$$V_{\min} \leq V_{(i,t)} \leq V_{\max} \qquad \forall i,t \quad (8)$$

$$I_{(l,t)} \leq I_{(l)}^{CAP} \qquad \forall l \in L, t \quad (9)$$

where $V_{\min}$ and $V_{\max}$ express the minimum and maximum voltage limits, i.e., 0.95 and 1.05 pu, respectively; $I_{(l,t)}$ demonstrate the per unit current through the line $l \in L$, and $I_{(l)}^{CAP}$ denote the current flow capacity in line $l$.

The total loading power is the summation of the consumption power by normal loads and PEVs:

$$P_{L(i,t)} = P_{NL(i,t)} + P_{L(i,t)}^{PEV} \quad (10)$$
$$\forall i \in I, t$$

$$Q_{L(i,t)} = Q_{NL(i,t)} + Q_{L(i,t)}^{PEV} \quad (11)$$
$$\forall i \in I, t$$

where $P_{L(i,t)}^{PEV}$ expresses the PEV active power at node $i$ and time $t$; $P_{NL(i,t)}$ and $Q_{NL(i,t)}$ represent the active and reactive powers of normal loads at node $i$ and time $t$, respectively. Because ac-dc inverter makes the dc link voltage constant, $P_{L(i,t)}^{PEV}$ is not dependent on the grid side voltage. Thus, reference [13] considers charging loads of PEV as a constant power in the load flow analysis. Another constraint for PEV is a restriction on injected reactive power. Injected reactive power from the PEV should be restricted by their converter nominal and dc-link voltages as shown below:

$$\left(Q_{L(i,t)}^{PEV}\right)^2 \leq \left(S_{L(i)}^{PEV}\right)^2 - \left(P_{L(i,t)}^{PEV}\right)^2 \quad (12)$$
$$\forall i \in I_{PEV}, t$$

$$\left(Q_{L(i,t)}^{PEV} + \frac{V_{(i,t)}^2}{X_{(i)}}\right)^2 \leq \left(\frac{V_{C(i)}^{max} V_{(i,t)}}{X_{(i)}}\right)^2 - \left(P_{L(i,t)}^{PEV}\right)^2 \quad (13)$$
$$\forall i \in I_{PEV}, t$$

where $S_{L(i)}^{PEV}$ represents the rated power of the PEV converter.

*3.1 Objective Function*

The purpose of this paper that is an optimal placement for PEVs to minimize voltages deviation, network power losses and the price of PEV based on constraints discussing in the previous part. Thus, the objective function can be obtained with considering the paper's aim and problem's constraints.

$$min : a.\sum_{t=1}^{T}\sum_{i=1}^{I}\left|V_i^{ref} - V_{(i,t)}\right| + b.\sum_{t=1}^{T} P_{Loss_t} + c.n_{PEV}.cost_{PEV} \quad (14)$$
$$s.t. \ (6)-(13)$$

where $T$ is the required period for analyzing the impact of all network's change on objective function; $n_{PEV}$ express the number of PEVs; $P_{Loss}$ indicates all losses in the power system; $V_i^{ref}$ is optimal voltage at the $i$th bus; and a, b, c are the weight coefficient. These factors are used to convert the multi-objective function to single-objective function.

## IV. CASE STUDY SIMULATIONS AND RESULTS

Figure 3 shows a 33 buses standard distribution network [14] which is simulated to validate the proposed method. In this paper, it is assumed that state estimation of distribution system has been done in the control center for the one-hour intervals by using the proposed algorithm in [15]. The weighted coefficients a, b, and c are equal to 0.6, 0.1, and 0.3, respectively. Furthermore, it is assumed that three (photovoltaic) PVs have been already installed on buses 7, 16, 33. These selections have been made due to the significant importance of the voltage profile in compared with other criteria. The simulation has been done for 24 hours. The optimal placement of PEVs parking lots has been done after installing PVs at specified buses. The optimal place for establishing PEVs which have been suggested by NSGA II and the proposed algorithm are nodes 19 and 32. Here, the distribution grid has been studied for three different conditions: neither PVs nor PEVs parking lots do not have reactive power exchange with the grid, only PVs can exchange reactive power with the power grid, both PVs and PEVs parking lots have the ability to exchange reactive power with the grid to improve the voltage profile.

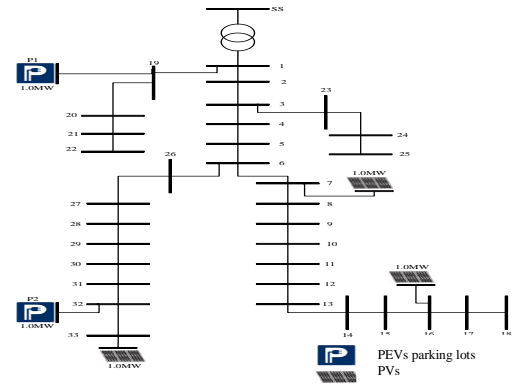

Fig. 3. A 33 buses standard distribution network with charging stations and PVs

Fig. 4 which is related to the sample voltage at the 10[th] bus for twenty-four hours shows that voltage profile is improved by injection of power to the distribution network by DGQ and V2GQ the.

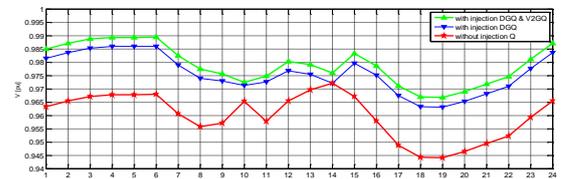

Fig. 4. sample voltage at the tenth bus for twenty-four hours

Fig. 5 represents all nodes' voltage for a particular hour. As can be seen, the voltage of all buses are improved by using DGQ and V2GQ.

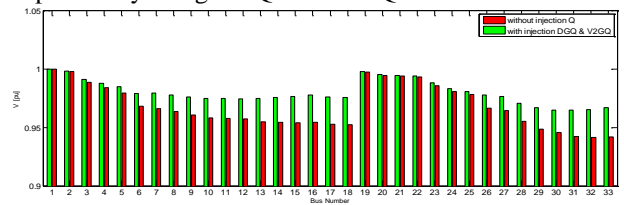

Fig. 5. All nodes' voltage for a particular hour

Fig. 6 shows reactive power exchange among PEVs parking lots at nodes 19 and 32, and distribution grid for 24 hours. It should be considered that the main goal of charge stations is providing the required active power pf PEVs, therefore, the capacity of these charging stations will be occupied as much as the PEVs needs active power and the remain will be used to generate/consume reactive power. In Fig. 6, positive values show the consumption of reactive power and the negative one shows the generation of reactive power.

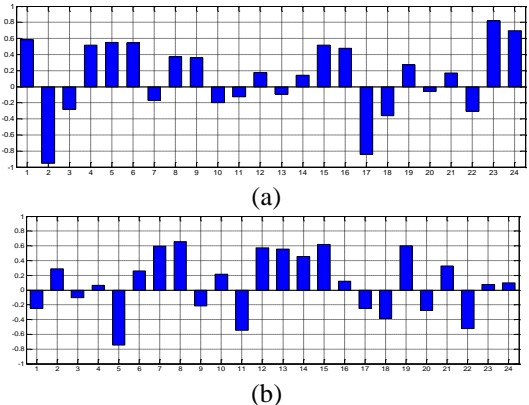

(a)

(b)

Fig. 6. Reactive power exchange at nodes 19(a) and 32(b)

## V. CONCLUSION

In this paper, a new method for optimal placement of PEVs parking lots by using the novel technology of V2GQ is presented. The results of implementing this approach for sample network confirm the fact that this method can be effective in a radial distribution grid. Since maintaining the bus voltage in a specified range is essential for the distribution system, this approach can be useful for improving voltage profile and keeping bus voltage in a specified range. Moreover, this technology can increase the profit from investment due to generate reactive power by PEVs. As a result, it can increase the persuasion of financier for financing in this area.

REFERENCES

[1] M. A. Azzouz, M. F. Shaaban, and E. F. El-Saadany, "Real-time optimal voltage regulation for distribution networks incorporating high penetration of PEVs," *IEEE Transactions on Power Systems,* vol. 30, pp. 3234-3245, 2015.

[2] M. Bollen and A. Sannino, "Voltage control with inverter-based distributed generation," *IEEE transactions on Power Delivery,* vol. 20, pp. 519-520, 2005.

[3] S. Deilami, A. S. Masoum, P. S. Moses, and M. A. Masoum, "Real-time coordination of plug-in electric vehicle charging in smart grids to minimize power losses and improve voltage profile," *IEEE Transactions on Smart Grid,* vol. 2, pp. 456-467, 2011.

[4] E. R. Joy, K. Thirugnanam, M. Singh, and P. Kumar, "Distributed active and reactive power transfer for voltage regulation using V2G system," in *Electric Power and Energy Conversion Systems (EPECS), 2015 4th International Conference on*, 2015, pp. 1-6.

[5] A. R. Malekpour, A. Pahwa, and B. Natarajan, "Distributed volt/var control in unbalanced distribution systems with distributed generation," in *2014 IEEE Symposium on Computational Intelligence Applications in Smart Grid (CIASG)*, 2014, pp. 1-6.

[6] T. Senjyu, Y. Miyazato, A. Yona, N. Urasaki, and T. Funabashi, "Optimal distribution voltage control and coordination with distributed generation," *IEEE Transactions on Power Delivery,* vol. 23, pp. 1236-1242, 2008.

[7] B. B. Zad, H. Hasanvand, J. Lobry, and F. Vallée, "Optimal reactive power control of DGs for voltage regulation of MV distribution systems using sensitivity analysis method and PSO algorithm," *International Journal of Electrical Power & Energy Systems,* vol. 68, pp. 52-60, 2015.

[8] A. Zakariazadeh, O. Homaee, S. Jadid, and P. Siano, "A new approach for real time voltage control using demand response in an automated distribution system," *Applied Energy,* vol. 117, pp. 157-166, 2014.

[9] M. Oshiro, K. Tanaka, A. Uehara, T. Senjyu, Y. Miyazato, A. Yona*, et al.*, "Optimal voltage control in distribution systems with coordination of distribution installations," *International Journal of Electrical Power & Energy Systems,* vol. 32, pp. 1125-1134, 2010.

[10] M. Degefa, M. Lehtonen, R. Millar, A. Alahäivälä, and E. Saarijärvi, "Optimal voltage control strategies for day-ahead active distribution network operation," *Electric Power Systems Research,* vol. 127, pp. 41-52, 2015.

[11] Rezvani, M. M., & Mehraeen, S. (2019). A New Approach for Steady-State Analysis of a Hybrid ac-dc Microgrid. arXiv preprint arXiv:1901.06011.

[12] J. Y. Yong, V. K. Ramachandaramurthy, K. M. Tan, and N. Mithulananthan, "Bi-directional electric vehicle fast charging station with novel reactive power compensation for voltage regulation," *International Journal of Electrical Power & Energy Systems,* vol. 64, pp. 300-310, 2015.

[13] M. Shaaban and E. El-Saadany, "Probabilistic modeling of PHEV charging load in distribution systems," in *Electric Power and Energy Conversion Systems (EPECS), 2013 3rd International Conference on*, 2013, pp. 1-6.

[14] S. Aslanzadeh, M. Kazeminejad, and A. Gorzin, "Comparision of voltage stability indicators in distribution systems," *Indian J. Sci. Res,* vol. 2, pp. 5-10, 2014.

[15] S. Naka, T. Genji, T. Yura, and Y. Fukuyama, "A hybrid particle swarm optimization for distribution state estimation," *IEEE Transactions on Power systems,* vol. 18, pp. 60-68, 2003.